\documentclass[journal=jacsat,manuscript=article]{achemso}

\usepackage{chemformula} 
\usepackage[T1]{fontenc} 

\usepackage{mathptmx}      
\usepackage{mhchem}

\usepackage{subfig}
\usepackage[decimalsymbol=point,range-units=single]{siunitx}
\usepackage{booktabs}
\usepackage[english]{babel}
\usepackage{xcolor}

\author{Cassia Lux}

\author{Thomas Tilger}
\author{Ramsia Geisler}
\author{Olaf Soltwedel}
\author{Regine von Klitzing}
\email{klitzing@smi.tu-darmstadt.de}
\phone{+49(0)6151-16-24506}
\affiliation[Unknown University]
{Soft Matter at Interfaces, Department of Physics, Hochschulstraße 8, 64289 Darmstadt, Germany}

\title[An \textsf{achemso} demo]
  {Model surfaces for paper fibers prepared from carboxymethyl cellulose and polycations}

\abbreviations{IR,NMR,UV}

\begin{document}


\begin{abstract}
\noindent For a tailored functionalization of cellulose based papers, the interaction between paper fibers and functional additives have to be understood. Planar cellulose surfaces present a suitable model system for studying the binding of additives. In this work, polyelectrolyte multilayers (PEMs) as model surfaces are prepared by alternating dip coating of the negatively charged cellulose derivate carboxymethyl cellulose and a polycation, either  PDADMAC or chitosan. The varied parameters of the PEM formation are the polyelectrolyte concentrations and pH (pH=2-6). Both PEM systems exhibit an exponential growth, which reveals a high mobility of the polyelectrolytes (PEs). The pH-tunable charge density leads to PEMs with different surface topographies. QCM-D experiments reveal pronounced viscoelastic properties of the PEMs. Ellipsometry and atomic force microscopy measurements show that the strong and highly charged polycation PDADMAC leads to the formation of smooth PEMs. The weak polycation chitosan results in cellulose model surfaces with higher film thicknesses and a tunable roughness. The PEMs prepared from both polycations exhibit a high water uptake when exposed to a humid environment. The resulting PEMs are suitable water-stable but water swellable model surfaces with a controllable roughness and topography.

\end{abstract}


\section{Introduction}
\label{intro}
\noindent Climate change has significantly increased the demand for recyclable and tunable materials with natural origin in the last years.\cite{Dufresne2008} A material that shows great promise in replacing plastics is cellulose based paper. In its unmodified state, paper is highly biodegradable, -compatible and recyclable.\cite{Tavakolian2020,Tang2019,Rinaudo2008} 
The cellulose fibers forming the network of paper are stabilized by hydrogen bonds and van-der-Waals forces. When the fibers get in contact with water, their swelling leads to a partial breakage of the inter-fibre stabilizing bonds.\cite{Dunlop-Jones1991} To make paper (more) stable towards water, it is a common procedure to use functional additives in the paper making process, which link the fibers through more stable inter-fibre connections via chemical or physical bonding.\cite{Gulsoy2014} An established method for the modification is the use of polymer resins such as amine epichlorohydrins and urea formaldehyde. These resins lead to a homo- and hetero-crosslinking of the resin and the cellulose fibers.\cite{Lindstrom2005,Rojas2011} The tuning of paper cannot only lead to a higher wet-strength, but can also modify characteristics such as flexibility and dry strength.\cite{Xu2004}\\

\noindent Planar cellulose model surfaces are a valid approach for investigating the functionalization of paper.\cite{Gunnars2002,Kontturi2006} Model surfaces allow to reduce measurement errors originating in the randomness of the fiber structure and open the possibility for many analysis methods for planar interfaces.\cite{Kontturi2006} In addition, by tuning the surface structure of the thin film, both the chemical and the physical factors of the interaction between the cellulose and a functional additive can be studied. Established methods for the preparation of model surfaces and thin films are spin-, dip-, and spray-coating and the Langmuir-Blodgett (LB) or Langmuir-Schäfer (LS) deposition methods.\cite{Kontturi2019} A general challenge in studying paper chemistry using model surfaces based on cellulose is the preparation of the model surface itself. Cellulose is not soluble in water and in most organic solvents due to the complex structure of the fiber\cite{Bismarck2002,Chinga2009} and the hydrogen bonds of the linear high molecular weight polysaccharide.\cite{Medronho2012} Exceptions are two-component-solvents, such as DMA/LiCl\cite{Dawsey1990}, and ionic liquids.\cite{Kargl2015} Successful dip- and spin-coating of cellulose containing thin films requires either a solution or a stable and homogeneous suspension prepared by dispersing cellulose colloids with dimensions in the size range of few nanometers. Edgar \textit{et al}. showed that cellulose nanocrystals in a stable aqueous suspension can be spin coated onto a substrate, resulting in a homogenous film with low roughness.\cite{Edgar2003} Using a cellulose solution with the aforementioned two-component solvents leads to the integration of a solid component into the thin film. Removing this component requires an additional washing step, which can alter the morphology of the model surface.\cite{Kontturi2019,Aulin2009}\\

\begin{figure}[H]
                 \centering
                 \includegraphics[clip,width=0.8\textwidth]{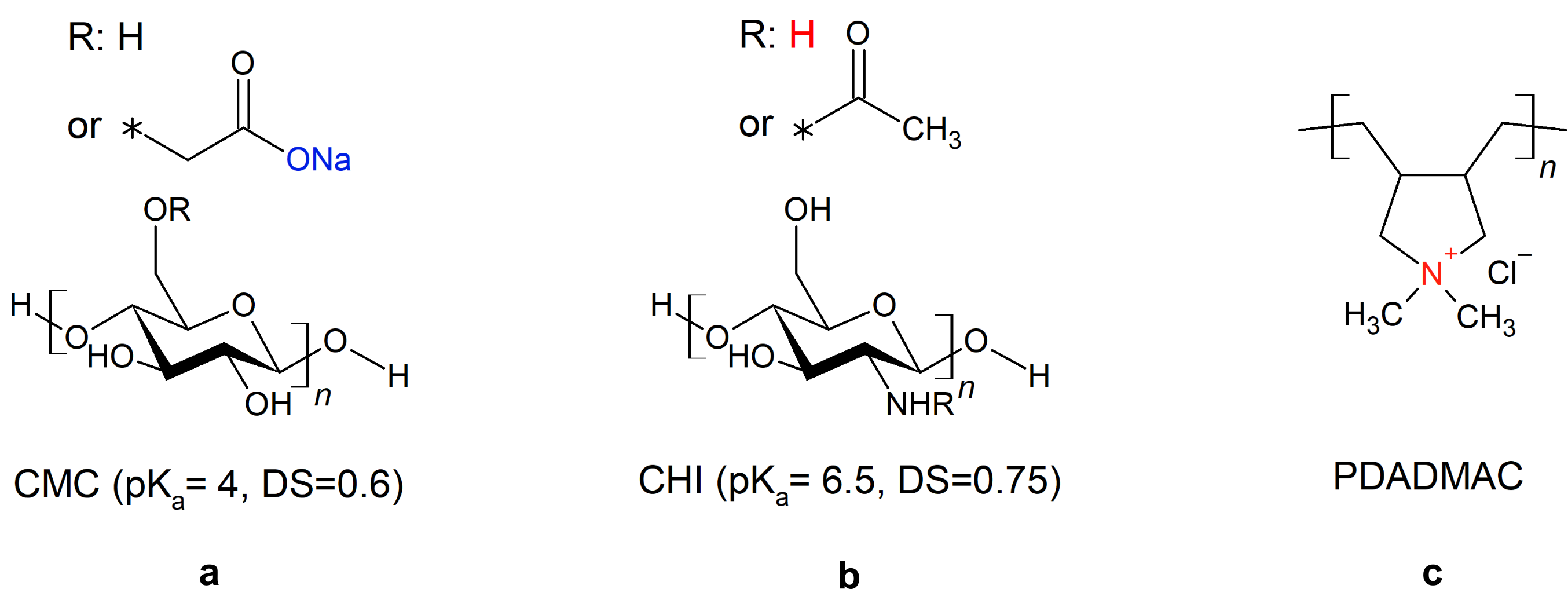}
                 \caption{Chemical structures of the used PEs. The negatively charged group of CMC (a) is depicted in blue, the positive charged groups of CHI (b) and PDADMAC (c) in red. For the weak PEs the  pK$_a$-value and the degree of substitution (DS) are given.}
                 \label{ChemStruc}
\end{figure}

\noindent In this work, cellulose model surfaces were prepared from an aqueous solution of carboxymethyl cellulose (CMC, Figure \ref{ChemStruc}a) by dip-coating. Compared to cellulose, in CMC the primary hydroxy groups at the C6-atoms are partly substituted with a carboxymethyl group. The negative charge of the carboxy group allows the formation of polyelectrolyte multilayers (PEM) by alternating dip-coating of CMC and a polycation. Chitosan (CHI, Figure \ref{ChemStruc}b) and Polydiallyldimethylammonium chloride (PDADMAC, Figure \ref{ChemStruc}c) are used as polycations in this study. The molecular structure of the biopolymer CHI is similar to that of cellulose, as it is a $\beta$-1,4 linked polyamino-saccharide. It can be used to tune both the wet- and dry-strength of paper through a physical linking of the fibers, as shown by Jahan \textit{et al}.\cite{Jahan2009} PDADMAC, on the other hand, is a strong synthetic polyelectrolyte (PE). The charge density of the weak PEs CMC (pK$_a \approx 4$) and CHI (pK$_a \approx 6.5$) are pH-dependent, while PDADMAC has a permanent positive charge. 

\noindent The goal of this work is to determine the influence of the chemical composition (PDADMAC vs. CHI) and preparation condition (pH, PE concentrations, CMC vs. polycation as outermost layer) on the morphorphology and swellability of the resulting PEMs. Besides the chemical structure, PDADMAC and CHI differ in their mobility and charge density. The system CHI/CMC prepared by dip-coating was partly subject of works by the groups of Beppu \textit{et al}.\cite{Taketa2018,Bataglioli2019,Spera2017} and Lui \textit{et al}.\cite{Zhang2013}. Beppu \textit{et al}. determined the influence of pH-value, salt concentration and molecular weight on the formation of CMC/CHI PEMs. Contrarily to their work, the prepared PEMs in this study are mainly terminated with the CMC layer. In addition to this, the outcome of the CHI/CMC PEMs are compared to the PDADMAC/CMC PEMs and the experimental work is extended by analytical methods such as Quartz crystal microbalance with dissipation monitoring (QCM-D) and ellipsometry to study the deposition kinetics and the swellability of the PEMs. The topography of the PEMs is visualized by AFM.

\section{Materials and methods}
\label{sec:exppart}

\subsection{Materials}
\label{materials}
\noindent The PEs sodium carboxymethyl cellulose (CMC, M\textsubscript{W}=\si{250} {kDa}, DS=0.6), polyethylenimine (PEI, M\textsubscript{W}= \si{750} {kDa}, 50 wt-\% in \ce{H2O}) and chitosan (CHI, M\textsubscript{W}=\si{50-190} {kDa}, DS=0.75) were purchased from Sigma Aldrich (Darmstadt, Germany). Linear polydiallyldimethylammonium chloride (PDADMAC, M\textsubscript{W}=\si{150} {kDa}) was purchased from PSS (Mainz, Germany). The ammonia solution (25 \%), hydrochlorid acid (\si{1} {mol/l}) and sodium hydroxide solution (\si{1} {mol/l}) were purchased from Merck KGaA (Darmstadt, Germany). Sulfuric acid (100\%) and hydrogen peroxide (30\%) were purchased from Carl Roth (Karlsruhe, Germany) and the acetic acid from VWR (Darmstadt, Germany). Ultrapure water was obtained from a Milli-Q-system from Merck with a resistance of \si{18} {M$\Omega$} cm.

\subsection{Preparation of PEMs}
\label{prepPEM}
\noindent The PEMs were prepared by dip-coating. As precursor, polyethylenimine was used at a concentration of \si{10^{-2}} {monoM}. The other PEs were dissolved in ultrapure water in the intended concentration of \SI{1}{g/l}, unless stated otherwise. For a better solubility of the polycation, \SI{0.1}{M} glacial acetic acid (HAc) was added to the CHI solution. For varying CHI concentration, the amount of HAc was adapted, so that the ratio of CHI and HAc remains constant (17.2 equiv. of HAc).  Using hydrochlorid acid and sodium  hydroxide solutions (0.1 or 1 M), the pH value of all PE solutions (excluding PEI) and the rinsing water was adjusted to 4, unless stated otherwise.\\

\noindent Double side polished silica wafers from Siegert Wafer (Aachen, Germany) were etched in piranha solution (3:1 with \ce{H2SO4} : \ce{H2O2}) for \SI{30}{min} and stored in ultrapure water until used but at most for \SI{2}{h}. Using a layer-by-layer dip robot (Riegler \& Kirstein, Berlin, Germany), the PEMs were prepared by first adsorbing the precursor PEI for \SI{30}{min}, followed by \SI{10}{min} of adsorption of the polyanion CMC and \SI{10}{min} of adsorption of the polycation CHI or PDADMAC. After each adsorption step, the layers were rinsed by dipping three times into rinsing water for \SI{2}{min}, \SI{1}{min}, and \SI{1}{min}. Through alternation of the subsequent adsorption of polyanion and polycation, the PEM forms with the desired number of bilayers (NoBL). In the present work, the PEI/CMC layer is defined as the first bilayer, therefore the top layer in a bilayer is CMC. 

\subsection{Ellipsometry}
\label{ellipsometry}

\noindent The measurement of the thickness and refractive index was carried out using a Null-Ellipsometer from Optrel (Sinzing, Germany) with a PCSA setup (polarizer-compensator-sample-analyzer). The wavelength of the laser was \SI{632.8}{nm}, the angle of incidence was set to \SI{70}{\degree}. The measurements for determining the swelling ratio of the PEMs (section \ref{Resswelling}) were carried out in a home-built humidity cell, which was placed in the beam path and connected to a nitrogen flow. By partly directing the nitrogen flow through two washing bottles filled with water at \SI{25}{\celsius}, the humidity inside the cell was controlled. The relative humidity and the temperature was measured by a thermohygrometer from Rotronic (Ettlingen, Germany). The ratio between dried and saturated nitrogen was adjusted to reach values of the relative humidity between 0 and \SI{95}{\%} RH, with at least 10 steps for each measurement. After reaching a constant reading of the humidity, the PEM was left to equilibrate until no change in the ellipsometric angles was observed anymore. When preparing a PEM on a silica wafer and measuring at ambient conditions, a two-layer model is required, as described in Table \ref{tab:layermodel}. The \ce{SiO_x} is a thin oxide layer on the substrate. As thickness of the oxide layer an average value was taken, which was determined by Löhmann \textit{et al}.\cite{Lohmann2018} by X-ray reflectometry.
\begin{table}[htbp]
  \centering
  \caption{Summary of the parameters for the two-layer model required for the analysis of the ellipsometric data.}
    \begin{tabular}{cccc}
    \toprule
    \textbf{layer} & \textbf{thickness / nm} & \textbf{n}     &\textbf{k} \\
    \midrule
    (humid) air   & continuum & 1.0000 & 0 \\
    PEM   & to be fitted & to be fitted & 0 \\
    SiOx  & 1.1   & 1.4570 & 0 \\
    Si    & continuum & 3.8858 & -0.0180 \\
    \bottomrule
    \end{tabular}%
  \label{tab:layermodel}%
\end{table}%
\subsection{AFM}
\label{afm}

\noindent Atomic force microscopy (AFM) was used to characterize the topography of the PEMs. The AFM measurements were carried out at ambient conditions ($\approx$\SI{40}{\%} RH) with the MFP-3D SA (Asylum Research, Oxford Instruments, California, USA). The cantilevers used are the AC160TS-R3 with a silicon probe and a tip diameter of 7-\SI{8}{nm}, also from Asylum Research. To determine the roughness of the PEMs, \SI{25}{\micro\meter ^2} images were taken and the roughness was determined from nine randomly selected \SI{1}{\micro\meter ^2}-areas of the image. The roughness is defined as the root mean square (RMS) of height deviations of the surface mean plane. For PEMs with surface patterns with a correlation length in the $\mu$m range, the roughness was determined on the whole \SI{25}{\micro\meter ^2} area. In all cases, the two different roughness measurements (\SI{1}{\micro\meter ^2}, \SI{25}{\micro\meter ^2}) lead to similar values. 

\subsection{QCM-D}
\label{qcm}
\noindent The adsorption kinetics of the PEM formation process is carried out in-situ by QCM-D measurements (QCM with dissipation monitoring Q-Sense Explorer, Biolin Scientific, Gothenburg, Sweden). The used crystals, coated with \ce{SiO2} and with a resonance frequency of \SI{4.95}{MHz}, were also purchased from Biolin Scientific. Prior to the measurement, the crystals were cleaned in an ultrasonic bath with chloroform, isopropanol and water, each for \SI{15}{min} and etched using the RCA procedure (5:1:1 of water:\ce{NH3}:\ce{H2O2} at \SI{75}{\celsius} for \SI{20}{min}). The baseline was recorded with the rinsing solution, the dipping order of the PE solutions was the same as in the corresponding dip-coating procedure. The time of adsorption and rinsing for each layer was the time  until the frequency reaches a steady state. The evaluation of the results was only done on the basis of the change in frequency and dissipation. The decrease in frequency is correlated with an increase in mass. An increase in dissipation indicates softening of the film. Comparing the different overtones of the signal gives additional information about the viscoelasticity of the thin film. Diverging overtones hint to a flexible and soft film.

\section{Results}

\noindent First, the PEMs are characterized in dependence of the preparation parameters NoBL, pH-value and PE concentration. In addition, in-situ QCM-D measurements are carried out and the change in frequency and dissipation is determined. Afterwards, the swellability of PEMs is determined in dependence of the varied parameters by measuring the change in thickness and refractive index with  varying relative humidity. 

\subsection{PEM formation}
\label{ResvariationBL}

\begin{figure*}[]
  \centering
  \includegraphics[width=0.95\textwidth]{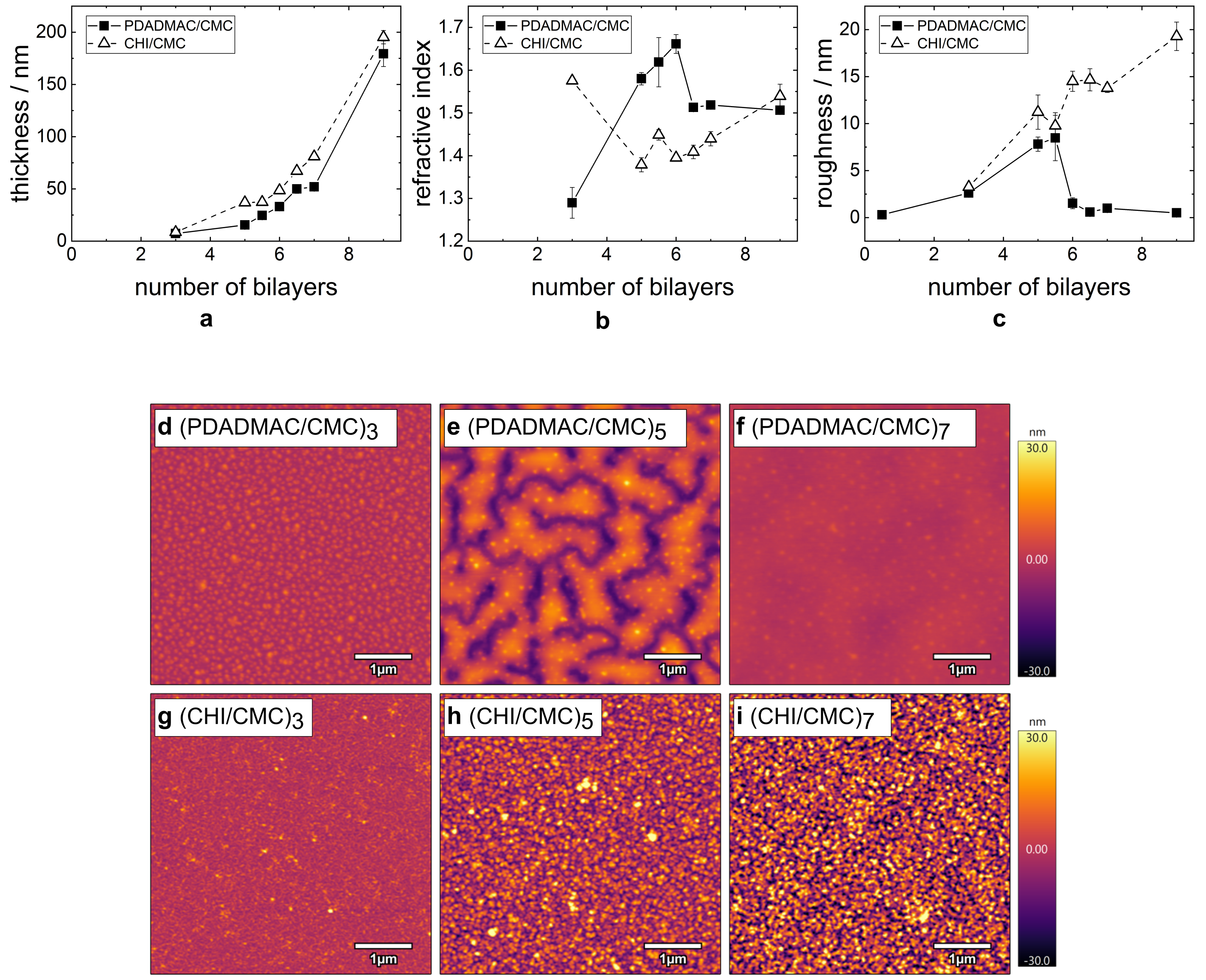}
  \caption{Summary of the morphology results for PEMS PDADMAC/CMC (filled squares) and CHI/CMC (empty triangles) from ellipsometry (a-b) and AFM measurements(c-i): (a) Film thickness, (b) refractive index and (c) roughness with an increasing NoBL. (d)-(i): AFM Images (\si{5x5}{\micro\meter $^2$}) for PDADMAC/CMC and CHI/CMC at 3, 5, and \SI{7}{BL}. For all images the height scale is set to \SI{60}{\nano\meter}. The samples were prepared at pH 4 and c$_PE$=\SI{1}{g/l}. All experiments were carried out at RH$\approx$40\%. The first bilayer corresponds to the bilayer of the precoat PEI and CMC.}
  \label{ElliAFMBL}
\end{figure*}

\noindent PEMs prepared from the polycations PDADMAC and CHI (pH=4; c$_{PE}$=\SI{1}{g/l}) with a varying NoBL are characterized with respect to their  thickness and morphology by ellipsometry (Figure \ref{ElliAFMBL} a-b) and AFM (Figure \ref{ElliAFMBL} c-i), respectively. In the figure, integers correspond to a bilayer terminating with the polyanion CMC and half numbers to the polycation as the outermost layer. The film thickness of both PE systems grows exponentially, whereas the CHI/CMC system results in slightly larger film thicknesses (Figure \ref{ElliAFMBL}a). The refractive index grows with increasing film thickness and NoBL, with the exception at \SI{3}{BL}, at which a much larger value is obtained (Figure \ref{ElliAFMBL}b). In contrast, the PDADMAC/CMC system shows a significant increase in the refractive index with increasing NoBL. At 6 BL, a plateau is reached leading to a constant value at higher BL. At 9 BL, the refractive indices of both PE systems are similar to each other. A refractive index above 1.6 is uncommon for PEMs, but will be discussed later in section \ref{Disformation}. \\

\noindent AFM measurements were carried out to determine the topography and the surface roughness of the PEMs. For selected NoBL, AFM images of PDADMAC/CMC and CHI/CMC are shown in Figures \ref{ElliAFMBL}d-i. In the case of PDADMAC/CMC at \SI{3}{BL} (Figure \ref{ElliAFMBL}d) an overall homogeneous PEM is obtained, which contains agglomerates in the size of few tens of nanometers. At \SI{5}{BL} (Figure \ref{ElliAFMBL}e), a pronounced domain formation is observed, and at \SI{7}{BL} (Figure \ref{ElliAFMBL}f), a closed and homogeneous film is obtained. The comparison of the measured thickness and the height scale of the topography shows that the sample surface at \SI{5}{BL} is not completely covered by PEM. The roughness (Figure \ref{ElliAFMBL}c) exhibits a maximum at 5.5 BL and then reaches a constant low value in the subnanometer range for PDADMAC/CMC.  The comparison of the PEMs PDADMAC/CMC (Figure \ref{ElliAFMBL}d) and CHI/CMC (Figure \ref{ElliAFMBL}g) at \SI{3}{BL} reveals that in both cases a homogenous surface is obtained, in the sense that there are no large agglomerates. With an increasing number of adsorbed bilayers, the surface roughness of CHI/CMC increases while the surface stays macroscopically homogeneous. Beyond \SI{6}{BL}, CHI/CMC exhibits a significantly higher roughness than PDADMAC/CMC.

\subsection{Adsorption kinetics}
\label{QCMvsBL}
\begin{figure*}[]
  \centering
\includegraphics[width=0.95\textwidth]{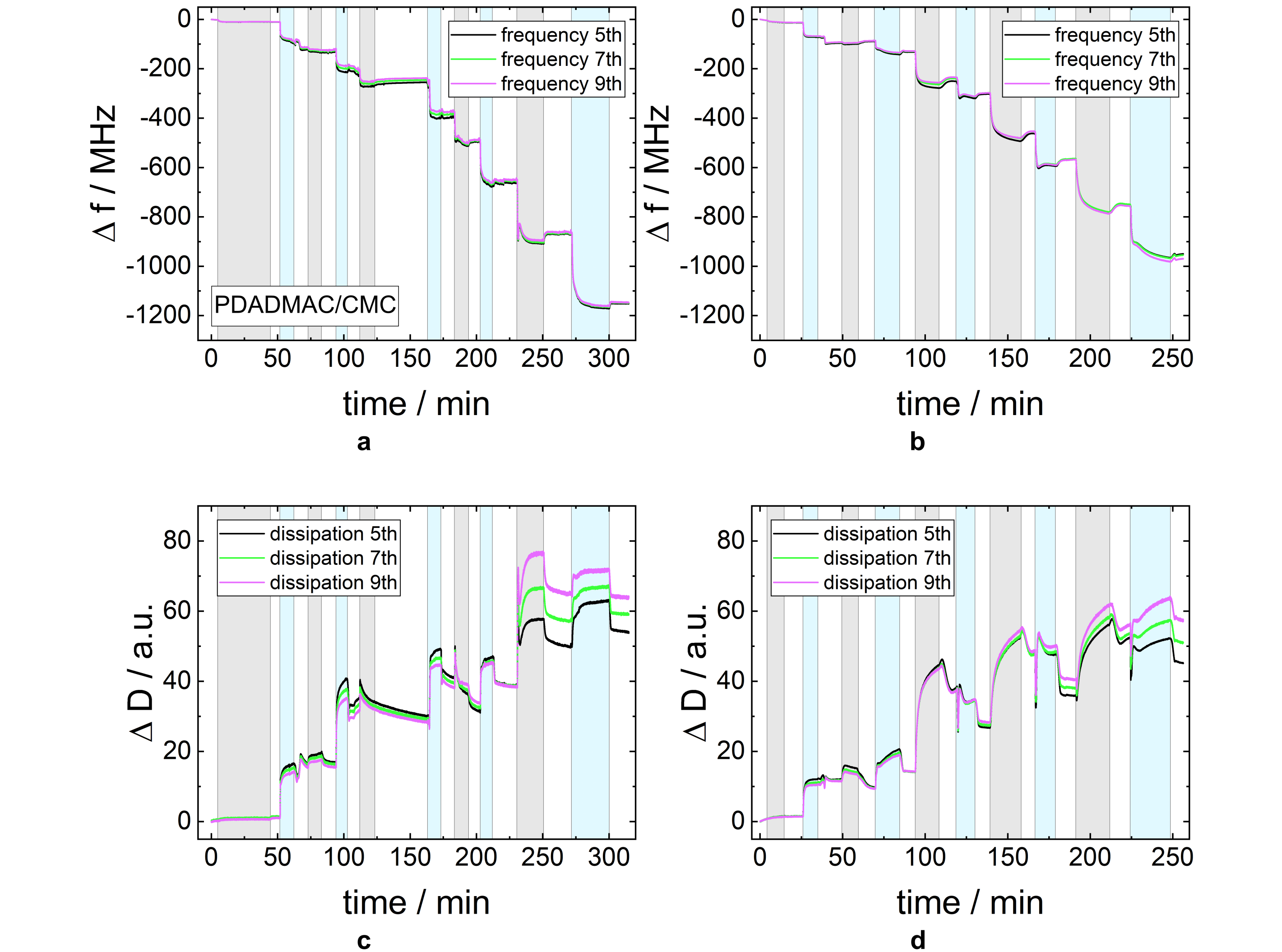}
  \caption{The change in frequency $\Delta$f (a-b) and dissipation $\Delta$D (c-d) measured by QCM-D for the PEMs PDADMAC/CMC and CHI/CMC. The gray region is the time period in which the polycation is adsorbed and the blue region in which the polyanion is adsorbed. The different curves in one plot represent the different overtones of the measured signal ( $5^{th}$, $7^{th}$, and $9^{th}$). All PE concentration were set to \SI{1}{g/l} and the pH-value was maintained at pH=4.}
  \label{1glQCMD}
\end{figure*}

\noindent To gain further insight into the adsorption behavior and to resolve the adsorption kinetics of the PEs, in-situ QCM-D measurements were carried out. Figure \ref{1glQCMD} shows the change in frequency and dissipation for PDADMAC/CMC and CHI/CMC. To facilitate the interpretation of the data, the adsorption time periods are highlighted by different colors: gray for the adsorption of the polycation, blue for the polyanion, white for the rinsing steps. The first area corresponds to the adsorption of the precursor layer PEI. Each adsorption step was carried out until the steady state is reached so that different adsorption times result. In the adsorption experiments of both PE systems, the overall frequency decreases, while the dissipation increases. The different overtones diverge with increasing NoBL. Each adsorption step can clearly be distinguished, as the contact of the charged layer with the PE solution induces an immediate change in frequency and dissipation. The time and the progress of frequency and dissipation until the steady state is reached varies for the two PEMs and for the different PEs in one system. It can also be observed that the amount of adsorbed mass and the time until steady state generally increases with each bilayer. The rinsing steps lead at most to a small amount of desorption and a small decrease in dissipation.\\    
          
\noindent Figure \ref{1glQCMD}a shows an increase in mass for each of the adsorption steps of CMC in the PDADMAC/CMC PEM system. The dissipation first increases significantly and then reaches a steady state value (Figure \ref{1glQCMD}c). The adsorption of CMC on the CHI layers follows a similar behavior in the first adsorption steps for the change in frequency and dissipation. For higher NoBLs, the change in dissipation during the adsorption of CMC exhibits a sharp decrease after the initial increase (Figure \ref{1glQCMD}d). The adsorption of CMC has a slightly higher impact on the change of both the frequency and the dissipation on the PDADMAC/CMC PEM than on the CHI/CMC PEM. The adsorption behavior of PDADMAC in the PDADMAC/CMC PEM differs noticeably from the expected growth behavior for the first layers. Both, the frequency and the dissipation decrease during adsorption. At a higher NoBL, the adsorption returns to the expected decrease in frequency and increase in dissipation. The adsorption of CHI in the CHI/CMC PEM also shows an increase in frequency and decrease in dissipation throughout each of the adsorption steps.

\subsection{Influence of pH-value}
\label{ResvariationpH}

\begin{figure}[]
  \centering
  \includegraphics[width=0.95\textwidth]{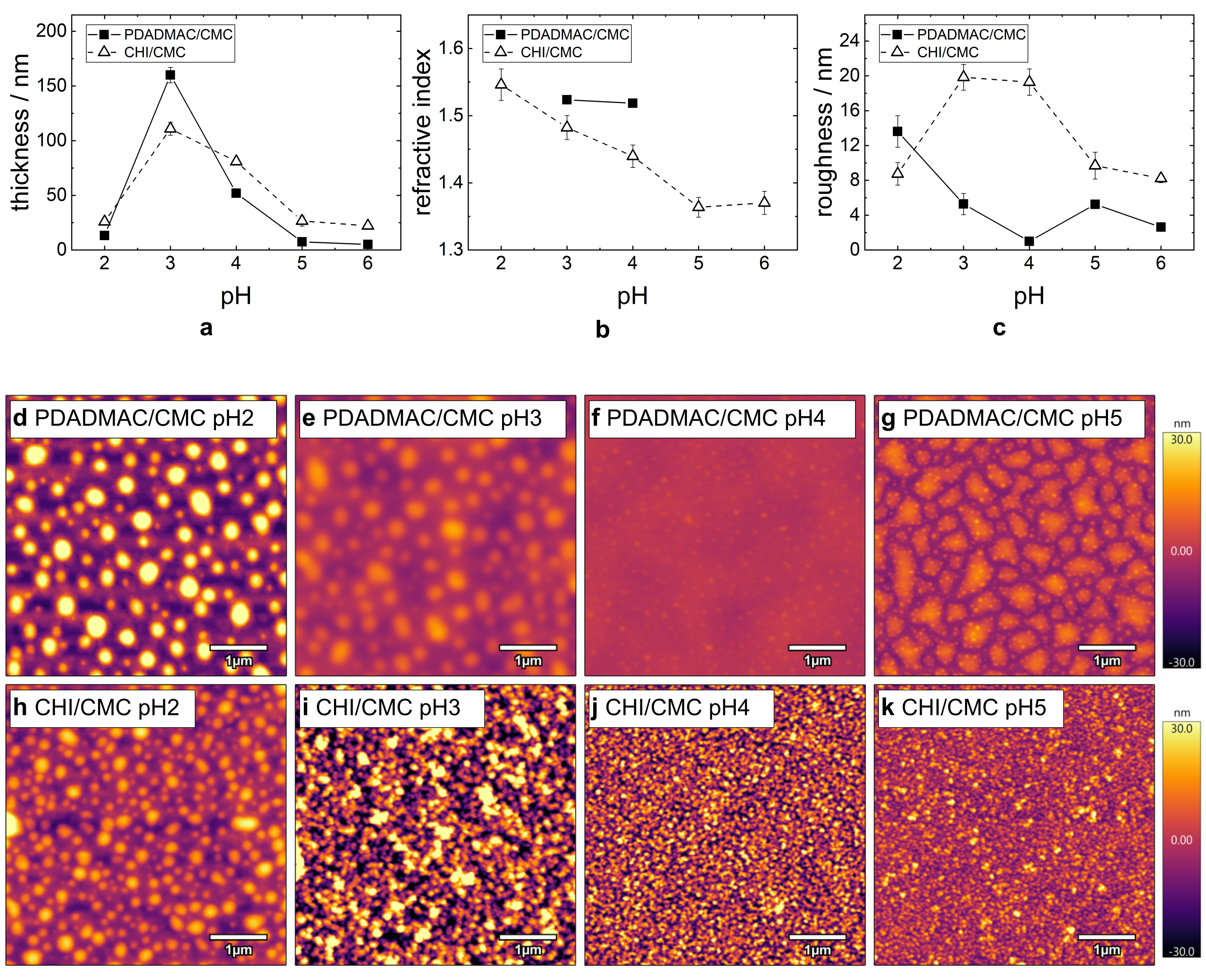}
  \caption{Summary of the morphology results for PEMS PDADMAC/CMC (filled squares) and CHI/CMC (empty triangles) from ellipsometry (a-b) and AFM measurements(c-i): (a) Film thickness, (b) refractive index and (c) roughness with varying pH-values. (d)-(i): AFM Images (\si{5x5}{\micro\meter $^2$}) for PDADMAC/CMC and CHI/CMC at pH=2, 3, 4 and 5. For all images the height scale is set to \SI{60}{\nano\meter}. The samples were prepared at c$_PE$=\SI{1}{g/l} and \SI{7}{BL}. All experiments were carried out at RH$\approx$40\%.}
  \label{ElliAFMpH}
\end{figure}

\noindent The respective thicknesses and refractive indices measured by ellipsometry for the PEMs (\SI{7}{BL}; c$_{PE}$=\SI{1}{g/l}) are shown in Figure \ref{ElliAFMpH}. Both, PDADMAC/CMC and CHI/CMC PEMs, exhibit maximum film thickness of the film at pH 3 (Figure \ref{ElliAFMpH}a). At pH 3, the film thickness of the PDADMAC/CMC PEM is larger than that for the CHI/CMC PEM, which is in contrast to all other pH-values. The refractive index decreases with increasing pH-values for the CHI/CMC PEM, reaching a constant value at about pH 5 (Figure \ref{ElliAFMpH}b). For PDADMAC/CMC, only the refractive indices for pH 3 and pH 4 are shown. Due to the small film thickness, the fitting of independent parameters (thickness and refractive index) was not reliable for the other pH-values. Nevertheless, it can be observed that the refractive indices of the PEMs at pH=3 and pH=4 for PDADMAC/CMC are similar.\\

\noindent The topography of PDADMAC/CMC and CHI/CMC PEMs in dependence of the pH and at \SI{7}{BL} was studied by AFM (Figure \ref{ElliAFMpH}d-k). Height images reveal for both PE systems  a strong agglomeration at pH 2 (Figure \ref{ElliAFMpH}d and h). The agglomerates are spherical and have lateral diameters of several hundreds of nanometers. As can be seen from the height and lateral scale, PDADMAC/CMC leads to larger agglomerates in all dimensions but a lower packing density compared to the CHI/CMC PEM. With increasing pH, the PDADMAC/CMC agglomerates decrease in size and the PEMs are overall smoother. The spherical aggregation progresses into a domain formation (Figure \ref{ElliAFMpH}g). Figure \ref{ElliAFMpH}c confirms the maximum roughness for pH 2 for PDADMAC/CMC and the smoothest PEM with the lowest roughness is obtained at pH 4. This is in agreement with the topography images, as large agglomerates correlate with a high roughness. For the CHI/CMC PEM the aggregate size decreases and a higher packing density with increasing pH-value is obtained (Figure \ref{ElliAFMpH}h-k). Here, the CHI/CMC PEMs have the largest roughness at pH 3 and pH 4 (Figure \ref{ElliAFMpH}c). Overall, the roughness is generally lower for the PDADMAC/CMC PEMs compared to the CHI/CMC PEMs. 

\subsection{Influence of PE concentration}
\label{ResvariationConcentration}
\begin{figure}[]
  \centering
  \includegraphics[width=0.95\textwidth]{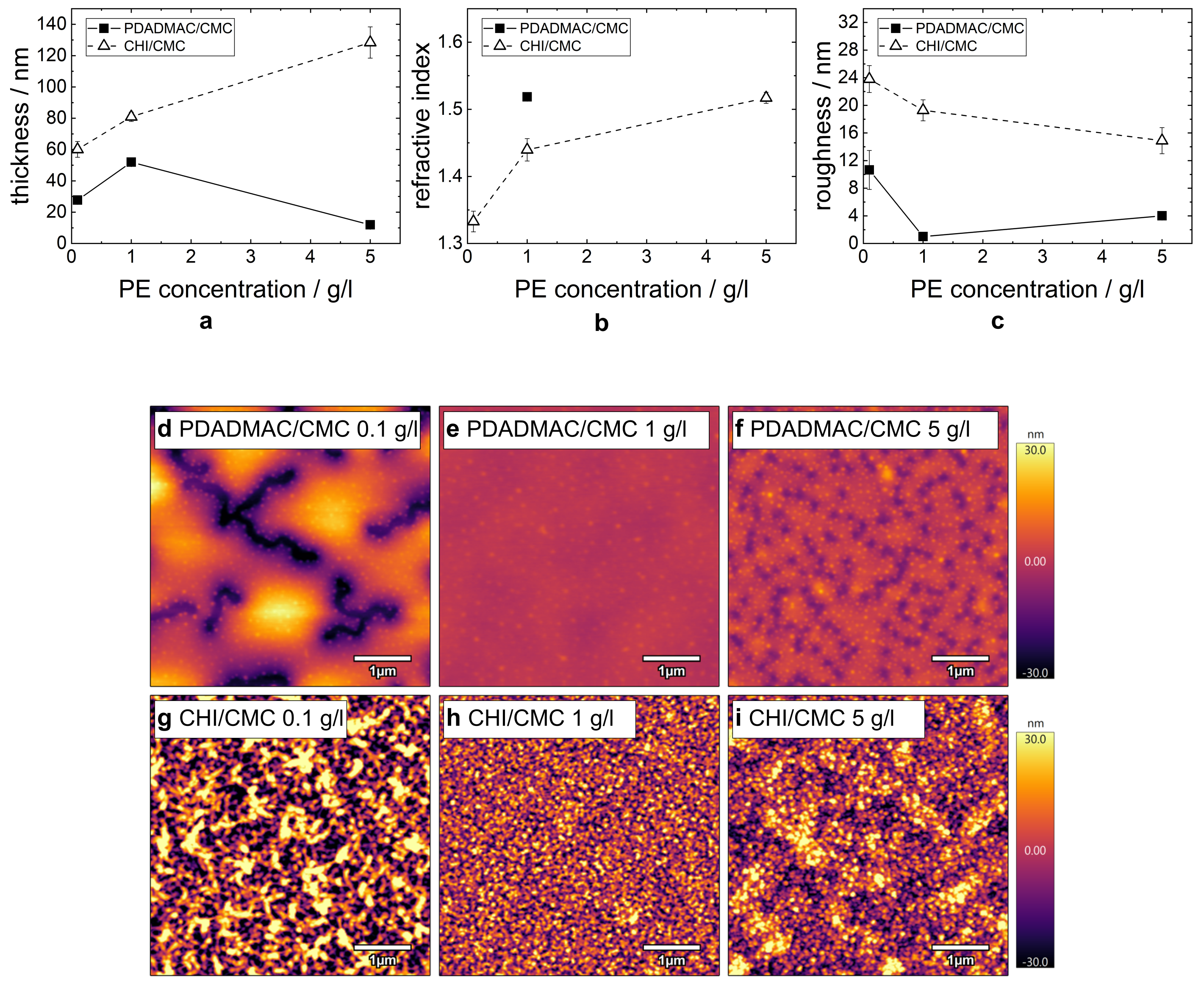}
  \caption{(a) and (b): Change in thickness and refractive index for PDADMAC/CMC (filled squares) and CHI/CMC (empty triangles) with varying PE concentration determined by ellipsometry (RH$\approx$40\%).  (c): Change in roughness, determined by AFM (RH$\approx$40\%), with varying PE concentration. for PDADMAC/CMC (filled squares) and CHI/CMC (empty triangles). (d)-(i): AFM Images (\si{5x5}{\micro\meter $^2$}) for PDADMAC/CMC and CHI/CMC varying PE concentration. For all images the height scale is set to \SI{60}{\nano\meter}. The NoBL was set to 7 and the concentration of all solutions was set to \SI{1}{g/l}. The first bilayer corresponds to the bilayer of the precoat PEI and CMC.}
  \label{ElliAFMConc}
\end{figure}

\noindent In the following, the influence of the PE concentration (0.1-\SI{5}{g/l}) on the thin film morphology is investigated. The respective morphology results of the prepared PEMs PDADMAC/CMC and CHI/CMC (\SI{7}{BL}; pH=4) are summarized in Figure \ref{ElliAFMConc} as function of the PE concentration. While the thickness of the PDADMAC/CMC PEMs is largest at a PE concentration of \SI{1}{g/l}, the thickness of the CHI/CMC PEMs linearly increases with the PE concentration within error bars  (Figure \ref{ElliAFMConc}a). For this system, the refactive index also increases with a rising PE concentration (Figure \ref{ElliAFMConc}b). For PDADMAC/CMC PEMs, the refractive index could only be determined for the thickest film prepared at \SI{1}{g/l}.\\

\noindent The PEM topography with varying PE concentration measured by AFM is shown in Figures \ref{ElliAFMConc}d-i. The PDADMAC/CMC PEMs prepared at 0.1 and 5 g/l shows a dewetting (Figures \ref{ElliAFMConc}d and f), while the PEM at 1 g/l is rather homogeneous (Figure \ref{ElliAFMConc}e). The domain formation at \SI{0.1}{g/l} is more prominent than for \SI{5}{g/l} and might lead to incomplete coverage. In agreement with the topography, the roughness for this system is lowest at \SI{1}{g/l} and highest at \SI{0.1}{g/l}. The surface of the CHI/CMC PEMs is more inhomogeneous of the PEMs prepared at \SI{0.1}{g/l} (Figure \ref{ElliAFMConc}g) and \SI{5}{g/l} (Figure \ref{ElliAFMConc}i) compared to the PEM prepared at \SI{1}{g/l} (Figure \ref{ElliAFMConc}h). The roughness decreases slightly with increasing concentration, which is in agreement with the topography images. Overall, the roughness of all CHI/CMC PEMs is significantly higher compared to the roughness of the PDADMAC/CMC PEMs.\\

\begin{figure*}[]
  \centering
\includegraphics[width=0.95\textwidth]{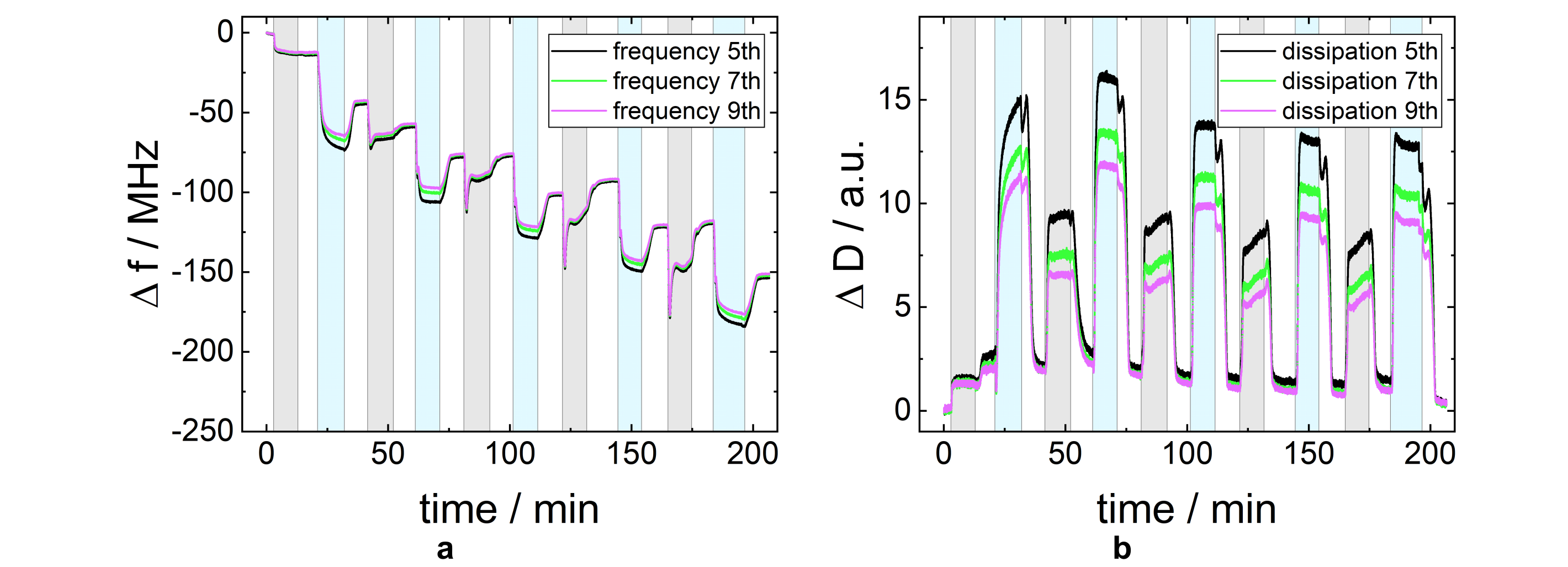}
  \caption{The change in frequency $\Delta$f (a) and dissipation $\Delta$D (b) measured by QCM-D for the PEM PDADMAC/CMC at \SI{5}{g/l}. The gray region is the time period in which the polycation is adsorbed, the blue region in which the polyanion is adsorbed and white for the rinsing periods. The different curves in one plot represent the different overtones of the measured signal ( $5^{th}$, $7^{th}$, and $9^{th}$). The pH-value was maintained at pH=4.}
  \label{5glQCM}
\end{figure*}

\noindent Since the low thickness of the PDADMAC/CMC PEM at \SI{5}{g/l} (Figure \ref{ElliAFMConc}a) was unexpected, the layer formation was additionally studied by QCM-D experiments (Figure \ref{5glQCM}). As the higher viscosity of the PE solutions impacts the shearing of the quartz crystal, a quantitative comparison between the different PE concentrations can only be done for the rinsing periods (Figure \ref{5glQCM}, white regions). The frequency change of the different rinsing periods shows a rather linear (instead of exponential) growth of the PEM. The decrease in frequency during the adsorption periods of CMC shows an increase in adsorbed mass  (Figure \ref{5glQCM}a, blue regions). In contrast to this, the adsorption of PDADMAC (Figure \ref{5glQCM}a, gray regions) leads to an initial increase in adsorbed mass, followed by slow increase in frequency, thus a desorption. The rinsing process after each PDADMAC adsorption step results in a final frequency equal to the frequency prior of the adsorption step. This means that the adsorption and rinsing of PDADMAC has no impact on the overall change in frequency for this layer and with that on the total adsorbed mass on the substrate. In addition, the resulting frequency change is significantly lower than for the PEMs prepared at \SI{1}{g/l} (Figure \ref{1glQCMD}a and c). Figure \ref{5glQCM}b shows that the film viscoelastic properties do not change significantly, as the dissipation only increases by a much slower rate, if at all, compared to the previously studied systems (Figure \ref{1glQCMD}b and d). The relatively stronger spreading of the overtones with respect to the overall dissipation during the adsorption steps is most likely an impact of the higher viscosity of the \SI{5}{g/l}. The results of the measurement with the QCM-D are in good agreement with the observed inhibited growth of the PDADMAC/CMC PEM prepared at \SI{5}{g/l} (Figure \ref{ElliAFMConc}a).

\subsection{Swelling of PEM}
\label{Resswelling}

\noindent PEMs are responsive materials, which respond to the surrounding relative humidity with a thickness change of the thin film. The swelling behavior of the PEMs is studied by measuring the thickness with ellipsometry at varying relative humidity (0 - 90\% RH). Through the increase of the relative humidity in the measuring chamber, the water content inside the PEM increases, leading to a swelling of the PEM. The swelling coefficient $S$ (Equation 1) being a measure for the water content, is calculated from the difference in thickness $d$ and the thickness of the dry PEM $d_0$ (1\% RH).

\begin{equation}
S=\frac{d-d_0}{d}.
\end{equation}

\begin{figure}[]
  \centering
\includegraphics[width=7in]{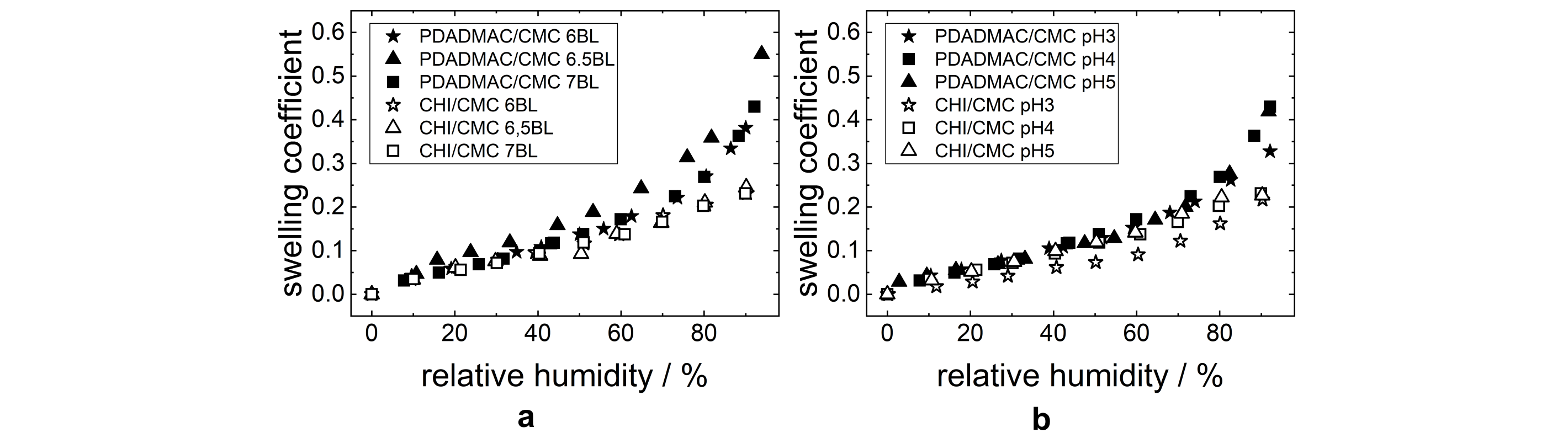}
\caption{The swelling coefficient $S$ in dependence of the relative humidity RH of the PEMs PDADMAC/CMC and CHI/CMC with (a) varying NoBL (at pH=4) and (b) different pH-values (\SI{7}{BL}). The thicknesses were measured by ellipsometry.}
\label{SwellCoef}
\end{figure}

\noindent Figure \ref{SwellCoef} shows the swelling coefficient as a function of RH for both investigated PE systems. With increasing humidity the film thicknesses of the studied PEMs PDADMAC/CMC and CHI/CMC increases (Figure \ref{SwellCoef}a). Nevertheless, a different swelling behavior is observed for the two PE systems. The PDADMAC/CMC PEMs first swell linearly up to 70\% RH. Above 70\% RH, the film thickness swells exponentially. In addition, the water uptake depends on the outermost PE layer (polycation or polyanion). As shown in figure \ref{SwellCoef}, the progression of the curves  for the PEMs with CMC as the outermost layer (\SI{6}{BL} and \SI{7}{BL}) is identical. For the PEM with PDADMAC as outermost layer (\SI{6.5}{BL}) a higher water uptake is observed. In the case of CHI/CMC, the swelling coefficient increases linearly with the relative humidity and is independent of the outermost PE. When comparing both PE systems to each other, it can be seen that the PDADMAC/CMC PEMs have overall higher swelling ratios, thus incorporate a higher quantity of water than the CHI/CMC PEMs.

\noindent The swelling behavior was also studied for PEMs prepared at different pH-values, as the charge density most likely has a structural effect on the thin film. Figure \ref{SwellCoef}b shows the swelling coefficient in dependence of the relative humidity and pH. The PDADMAC/CMC PEMs swell in a similar manner for all pH-values (3, 4 and 5). The CHI/CMC PEMs at pH 4 and 5 show a similar linear swelling behavior compared to the PEMs presented in Figure \ref{SwellCoef}a. The swelling behavior of the thin film prepared at pH 3 is lower than for the PEMs prepared at pH 4 and 5. An exponential upturn at around 70\% RH leads to superposition with the PEMs prepared at pH 4 and 5.

\section{Discussion}
\label{discussion}
\noindent This work compares the influence of PE concentration, pH and the NoBL on PEM formation composed of varying polycations (PDADMAC or CHI) and the polyanion (CMC) with respect to the layer formation properties and surface topography characteristics. In addition, the swelling behavior at varying relative humidity is demonstrated for exemplary samples with a varying NoBL and different pH. The main analytical methods are AFM, ellipsometry and QCM-D experiments. 

\noindent In the following, first, the PEM formation process and the surface morphology are discussed. Next, the influence of the structure on the swellabillity of the PEMs is elaborated.
 
\subsection{Multilayer formation}
\label{Disformation}

\noindent Both PE systems (PDADMAC/CMC and CHI/CMC) grow exponentially with an increasing NoBL (Figures \ref{ElliAFMBL}a, \ref{1glQCMD}a and \ref{1glQCMD}c). An exponential thickness growth of the PEM is linked to a high mobility of the PEs.\cite{Lavalle2004} The mobility of the PEs determines the ability for diffusion between the bulk PEM and the solvent. For example, PEs with a high charge density and no added salt, as is the case for PDADMAC (Figure \ref{ChemStruc}c), do not diffuse into the PEM. Systems with immobile PEs, for example PDADMAC/PSS, leads to a linear thickness increase for each adsorption step.\cite{Tang2016,Volodkin2014} From that it is concluded that the exponential growth observed for the PDADMAC/CMC PEM results from a high mobility of CMC (Figure \ref{ChemStruc}a). In the CHI/CMC system both PEs are mobile enough in order to diffuse in the PEM. While, the high mobility of CMC is deduced from the thickness results of the PDADMAC/CMC system, the mobility of CHI is already confirmed in literature for the PE system Hyaluronan/CHI.\cite{Kujawa2005} Since the high mobility of the PE contributes to the thickness growth, it is expected that the mobile and coiled CHI (Figure \ref{ChemStruc}b) contributes by a higher fraction to the thickness growth compared to the immobile and rather rigid PDADMAC molecules. This in turn means that the similar layer thicknesses (Figure \ref{ElliAFMBL}a and \ref{1glQCMD}a,c) for both PE systems point to a synergistic effect of PDADMAC and CMC contributing to the large resulting thickness of the film. During the adsorption step of PDADMAC, CMC is assumed to diffuse out of the PEM bulk to the surface leading to an intrinsic charge compensation.\cite{Volodkin2014,Lavalle2004,Abdelkebir2011,Porcel2006} This charge compensation in turn leads to a further PDADMAC adsorption. The pronounced adsorbed amount of PDADMAC is supported by a pronounced change in frequency (Figure \ref{1glQCMD}a). Since PDADMAC is the more rigid than CMC a pronounced decrease in dissipation occurs at the same time. The complexation of CMC and PDADMAC at the PEM surface and the release of counter ions leads to a denser layer and reduced layer softness (or dissipation).\cite{Wang2016}\\

\noindent The results of the pH variation on the PEM morphology is related to the PEM structure depending on the charge density of the PE (Figure \ref{ElliAFMpH}). On the one hand, a high charge density correlates with a slower PE diffusion and the adsorption of the stretched chains contributes less to the film thickness. On the other hand, a too low charge density can inhibit the formation of the PEMs. This is due to a lacking change in surface charge during the alternating adsorption of polycation and -anion.\cite{V.Klitzing2006,Steitz2001} This becomes visible in the ellipsometry measurements (Figure \ref{ElliAFMpH}a). PDADMAC is a strong PE carrying one permanent and pH-independent charge on each monomer (Figure \ref{ChemStruc}c). CMC is a weak PE (Figure \ref{ChemStruc}a, DS=0.6, pK$_\text{a}\approx$4), whose charge density varies with the protonation state and thus with the pH of the PE solution.  At low pH-values, the charge density of CMC is very low due to protonation of the carboxylic group. This  results in an inhibited PEM assembly, lower film thicknesses (Figure \ref{ElliAFMpH}a) and higher agglomeration (Figure \ref{ElliAFMpH}c). At pH 5 and 6 , the higher charge density leads to stretched chain conformation and small film thicknesses of the PEMs. At a pH-value between 3 and 4, a balance between the charge density and the coiling of the chains exists, resulting in the observed maximum film thickness (Figure \ref{ElliAFMpH}a). The effect of the PE solution pH on the film thickness is similar for the CHI/CMC as for the PDADMAC/CMC PEMs. This trend confirms the major effect of CMC on the pH-dependency of the PEMs. The charge density of CHI (DS=0.75, pK$_\text{a}\approx$6.5) does not vary between pH 2 and 6 due to almost complete protonation over the entire pH range. A pH-dependent effect of CHI is therefore not expected (see also Zhang et al. \cite{Zhang2013}) and was not observed during the experiments. At pH 3, a significantly higher film thickness is obtained for the PDADMAC/CMC PEM compared to the CHI/CMC PEM (Figure \ref{ElliAFMpH}a) due to the higher charge density of PDADMAC than for CHI. Here, it is assumed that the mobility of the weakly charged CMC is further enhanced by PDADMAC than by CHI and a higher amount of PDADMAC is adsorbed. \\

\noindent The effect of the PE concentration of the dipping solutions is shown in Figures \ref{ElliAFMConc} and \ref{5glQCM}. The formation of PEMs is strongly controlled by the adsorption rate of the PE to the surface of the PEM. Therefore, a decrease in PE concentration influences the amount of adsorbed PE.\cite{V.Klitzing2006} If the PE concentration is decreased but the adsorption time is maintained constant at \SI{10}{min}, a reduced film thickness is expected.\cite{Dubas1999,Garg2008,Mermut2003} This effect is confirmed for both PEM systems PDADMAC/CMC and CHI/CMC (Figure \ref{ElliAFMConc}a). Here, the lowest PE concentration leads to the thinnest PEMs in both cases. The higher roughness observed (Figure \ref{ElliAFMConc}c) may result from an insufficient surface coverage. As expected, an increasing PE concentration leads to a higher film thickness of the resulting PEMs (Figure \ref{ElliAFMConc}a). For PDADMAC/CMC, the PEM formation is not monotonous since a stripping of PE complexes leads to a partial desorption (Figures \ref{ElliAFMConc}a and \ref{5glQCM}). This effect was already observed by Sui \textit{et al}.\cite{Sui2003} and is confirmed by the results of this study.
\subsection{Structure and Swellability}
\label{Disstructure}
\noindent In the manuscript, it is demonstrated that the structure and morphology of the PEMs strongly depend on the preparation conditions such as NoBL, pH, and PE concentration of the dipping solution. The domain formation shown in the AFM images (Figures \ref{ElliAFMBL}d-f) is assumed to result from a lateral agglomeration during the drying of the PEM rather than by the rinsing steps\cite{Fery2001,Lehaf2012}. The formation of large domains explains the high refractive indices (Figure \ref{ElliAFMBL}b) and roughness (Figure \ref{ElliAFMBL}c) of the intermediate NoBL. The high roughness and its potential influence on the optical fitting of the ellipsommetric data can lead to an underestimated film thickness and overestimated refractive index.\cite{Fenstermaker1969,Meier2011} The dewetting of the PDADMAC/CMC PEMs hints to a flexible thin film. This is further supported by the low local roughness (Figure \ref{ElliAFMBL}c, after \SI{6}{BL}), which is independent of the NoBL. A roughness is independent of the film thickness when the PEMs are flexible enough to decrease interfacial tensions, thus the roughness of the PEM, or for PEM systems with highly mobile PEs diffusing into the layer.\cite{Lehaf2012} The high flexibility of the PDADMAC/CMC complexes facilitates its high water uptake (Figure \ref{SwellCoef}a). The PDADMAC terminated PEM (Figure\ref{SwellCoef}a, \SI{6.5}{BL}) shows a higher water uptake because of a higher concentration of counter ions\cite{Ghoussoub2018} and therefore a higher osmotic pressure in the PEM.\cite{Lohmann2018,Wong2004} This dependency of the water uptake on the outermost PE layer is commonly observed for PEMs and known as the odd-even effect.\cite{Wang2017,Zerball2015,Wang2016,Nestler2012,Wong2004}\\

\noindent The CHI/CMC system does not result in observable flexible PEMs. On the one hand, Figure \ref{ElliAFMBL}c shows an increasing roughness with increasing NoBL, as commonly observed for PEMs.\cite{Lavalle2002} This observation and the overall higher roughness compared to PDADMAC/CMC PEM indicates that the flexibility of the CHI/CMC PEM is not high enough in order to reduce interfacial tensions through a smoothing of the surface. On the other hand, the lower flexibility reduces the water uptake of the CHI/CMC PEMs (Figure \ref{SwellCoef}). The overall lower water uptake of the CHI/CMC PEMs is assumed to result from a lower concentration of counter ions. Moreover, no prominent odd-even effect is observed since similar swelling coefficients are obtained for the PEM with CHI and with CMC as outermost layer. However, the water uptake at pH 3 is lower compared to the PEMs prepared at the higher pH-values and increases exponentially above \SI{60}{\%} RH. The strong chain coiling and low charge density of CMC at pH 3 seems to lead to a denser film with less counter ions present in the film. The non-linear behavior beyond \SI{60}{\%} RH may result from an enhanced flexibility of the film at high humidity and the release of counter ions in the polymer coils. In conclusion, the higher flexibility of the PDADMAC/CMC PEMs is assumed to result from a higher concentration of counter ions generally present in the PEM. On the one hand, the higher concentration of counter ions presumably stems from a higher extrinsic charge compensation caused by the highly charged PDADMAC. On the other hand, the high mobility of CMC and CHI leads to a complexation of the two PEs and a decrease in counter ion concentration.\\ 

\section{Conclusion}
\label{conclusion}

\begin{figure*}[]
  \centering
\includegraphics[width=0.95\textwidth]{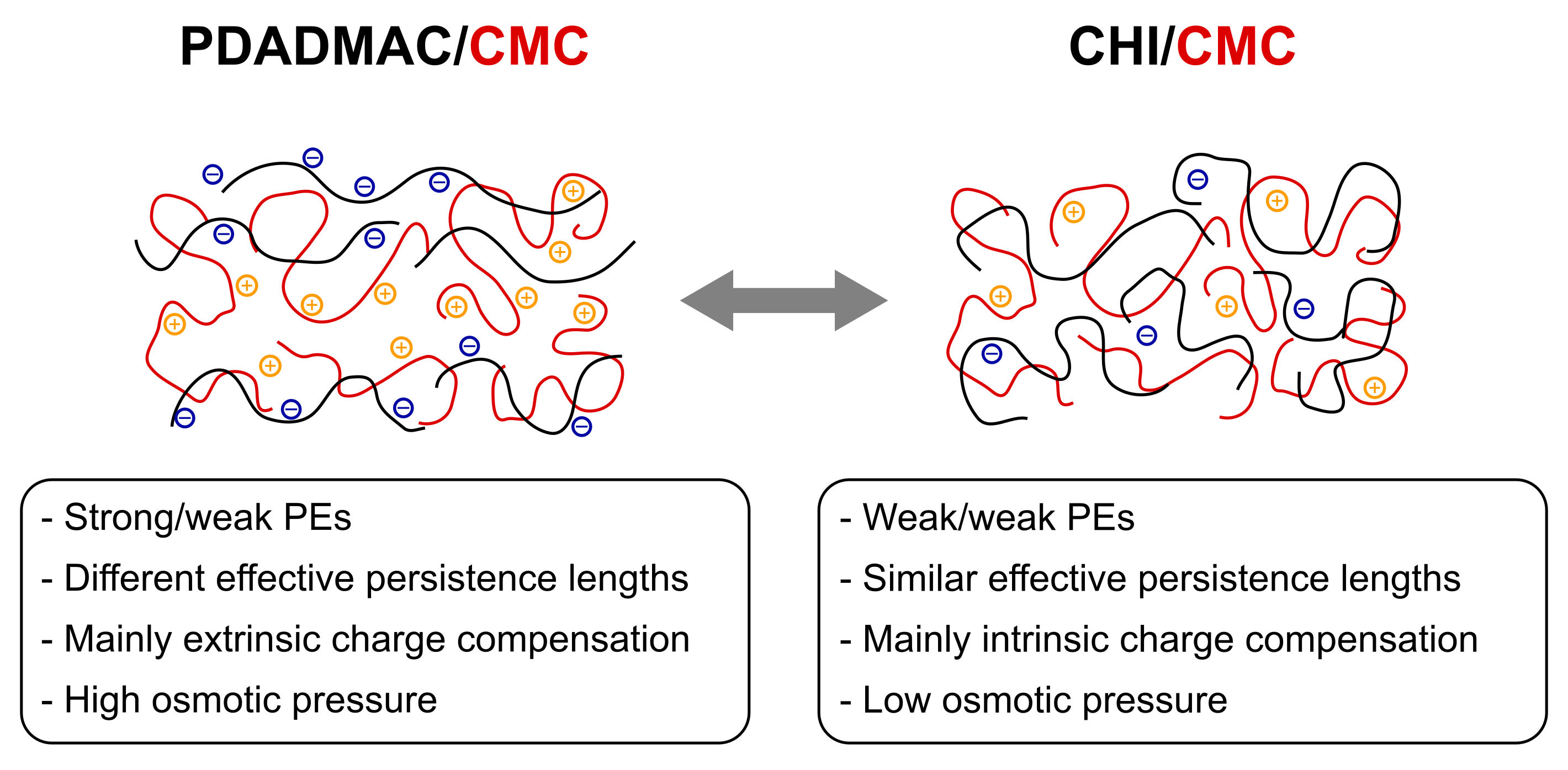}
  \caption{Proposed structures of the PEM systems PDADMAC/CMC and CHI/CMC, resulting from the high charge density of PDADMAC and the high effective persistence length. The polycations are depicted in black, the corresponding negatively charged counter ions in blue. The polyanions are depicted in red, the corresponding positively charged counter ions in orange.}
  \label{Scheme}
\end{figure*}

\noindent In this work, thin films in the form of PEMs are prepared by alternating dip-coating of the cellulose derivate CMC and the positively charged polylectrolytes PDADMAC or CHI. These PEMs enable the study of the interaction between cellulose fibers with functional additives used for the modification of paper products. By using the cellulose derivate CMC, the water insolubility of cellulose is overcome and the formation of PEMs enabled. These PEMs exhibit unique film properties such as a controllable thickness and surface topography, which were characterized with a broad variety of experimental methods. The study reveals that PEMs formed with PDADMAC and CMC are flexible thin films. The PEMs prepared with more than \SI{6}{BL} have a low roughness and show a homogeneous surface topography, which can be related to highly charged PDADMAC chains, which are adsorbed in a more stretched conformation than the CMC chains. 
Figure \ref{Scheme} shows a proposed structure for both PEM systems. Due to the high charge density, PDADMAC (Figure \ref{Scheme}, left) has a longer effective persistence length (includes backbone rigidity and charge effects) than CMC and CHI. Therefore, 1:1 stoechiometric complexation between PDADMAC and CMC is not fulfilled, which leads to pronounced extrinsic charge compensation by non-adsorbed counter ions in the PEM. The high amount of counter ions leads to a high osmotic pressure and a high water uptake of the PDADMAC/CMC PEMs. In addition, a lower amount of complexation sites allows parts of the PEM to move which might explain the ability to form lateral domains under certain conditions. The similarity of persistence lengths of CHI and CMC (originating from similar backbone and charge densities) leads mainly to intrinsic charge compensation and a low osmotic pressure of the CHI/CMC PEM (Figure \ref{Scheme}, right). Since both polysaccharides have bulky monomers they form quite large loops leading to larger agglomerates (roughness) and thickness than the PDADMAC / CMC PEM under certain pH conditions. \\

\noindent Cellulose model surfaces are used for the study of the functionalization of paper, as it not only enables the broadening of the range of possible analysis methods, but also excludes errors from the randomness and complex structure of the fiber surface. The preparation of PDADMAC/CMC and CHI/CMC PEMs should allow an extensive study of fiber-polymer interaction. The PDADMAC/CMC PEMs are thin and smooth, independent of the NoBL (above a thickness of \SI{50}{\nano\meter}). The smooth film enables the study of the chemical interaction of functional additive with the model surface. On the other hand, the surface of the CHI/CMC PEMs with controlled roughness can mimic the rough surface of a fiber and can be used to study the physical share of the interaction. It is known that the swelling of the fibers during the paper preparation has an influence on the modification and the studied swelling behavior of the PEMs can be used to further study the effect of the integration of the functional additives into the fiber wall on the functionalization.

\begin{acknowledgement}
The authors thank the Deutsche Forschungsgemeinschaft (DFG) under the grant PAK 962-1, subproject B4 (KL1165/28-1) for the funding of this project.

\end{acknowledgement}

\subsection{Conflict of Interest}
The authors have no conflict to declare.

\subsection{Author Contributions}
Conceptualization, C.L., R.v.K.; formal analysis, C.L., T.T.; investigation,  C.L., T.T.;  writing--original draft preparation, C.L.; writing--review and editing, C.L.,R.G.,O.S.,R.v.K.; visualization, C.L.; supervision, R.v.K.; project administration, R.v.K.; funding acquisition, R.v.K. All authors have read and agreed to the published version of the manuscript.


\bibliography{CMSCluxArXive}

\providecommand{\latin}[1]{#1}
\makeatletter
\providecommand{\doi}
  {\begingroup\let\do\@makeother\dospecials
  \catcode`\{=1 \catcode`\}=2 \doi@aux}
\providecommand{\doi@aux}[1]{\endgroup\texttt{#1}}
\makeatother
\providecommand*\mcitethebibliography{\thebibliography}
\csname @ifundefined\endcsname{endmcitethebibliography}
  {\let\endmcitethebibliography\endthebibliography}{}
\begin{mcitethebibliography}{49}
\providecommand*\natexlab[1]{#1}
\providecommand*\mciteSetBstSublistMode[1]{}
\providecommand*\mciteSetBstMaxWidthForm[2]{}
\providecommand*\mciteBstWouldAddEndPuncttrue
  {\def\EndOfBibitem{\unskip.}}
\providecommand*\mciteBstWouldAddEndPunctfalse
  {\let\EndOfBibitem\relax}
\providecommand*\mciteSetBstMidEndSepPunct[3]{}
\providecommand*\mciteSetBstSublistLabelBeginEnd[3]{}
\providecommand*\EndOfBibitem{}
\mciteSetBstSublistMode{f}
\mciteSetBstMaxWidthForm{subitem}{(\alph{mcitesubitemcount})}
\mciteSetBstSublistLabelBeginEnd
  {\mcitemaxwidthsubitemform\space}
  {\relax}
  {\relax}

\bibitem[Dufresne(2008)]{Dufresne2008}
Dufresne,~A. \emph{Monomers, Polymers and Composites from Renewable Resources};
  Elsevier, 2008; pp 401--418\relax
\mciteBstWouldAddEndPuncttrue
\mciteSetBstMidEndSepPunct{\mcitedefaultmidpunct}
{\mcitedefaultendpunct}{\mcitedefaultseppunct}\relax
\EndOfBibitem
\bibitem[Tavakolian \latin{et~al.}(2020)Tavakolian, Jafari, and van~de
  Ven]{Tavakolian2020}
Tavakolian,~M.; Jafari,~S.~M.; van~de Ven,~T. G.~M. {A Review on
  Surface-Functionalized Cellulosic Nanostructures as Biocompatible
  Antibacterial Materials}. \emph{Nano-Micro Letters} \textbf{2020}, \emph{12},
  73\relax
\mciteBstWouldAddEndPuncttrue
\mciteSetBstMidEndSepPunct{\mcitedefaultmidpunct}
{\mcitedefaultendpunct}{\mcitedefaultseppunct}\relax
\EndOfBibitem
\bibitem[Tang \latin{et~al.}(2019)Tang, Liu, Zhang, He, Li, Xu, Ni, and
  Li]{Tang2019}
Tang,~R.~H.; Liu,~L.~N.; Zhang,~S.~F.; He,~X.~C.; Li,~X.~J.; Xu,~F.; Ni,~Y.~H.;
  Li,~F. {A review on advances in methods for modification of paper supports
  for use in point-of-care testing}. \emph{Microchimica Acta} \textbf{2019},
  \emph{186}, 521\relax
\mciteBstWouldAddEndPuncttrue
\mciteSetBstMidEndSepPunct{\mcitedefaultmidpunct}
{\mcitedefaultendpunct}{\mcitedefaultseppunct}\relax
\EndOfBibitem
\bibitem[Rinaudo(2008)]{Rinaudo2008}
Rinaudo,~M. {Main properties and current applications of some polysaccharides
  as biomaterials}. \emph{Polymer International} \textbf{2008}, \emph{57},
  397--430\relax
\mciteBstWouldAddEndPuncttrue
\mciteSetBstMidEndSepPunct{\mcitedefaultmidpunct}
{\mcitedefaultendpunct}{\mcitedefaultseppunct}\relax
\EndOfBibitem
\bibitem[Dunlop-Jones(1991)]{Dunlop-Jones1991}
Dunlop-Jones,~N. \emph{Paper Chemistry}; Springer Netherlands: Dordrecht, 1991;
  pp 76--96\relax
\mciteBstWouldAddEndPuncttrue
\mciteSetBstMidEndSepPunct{\mcitedefaultmidpunct}
{\mcitedefaultendpunct}{\mcitedefaultseppunct}\relax
\EndOfBibitem
\bibitem[Gulsoy(2014)]{Gulsoy2014}
Gulsoy,~S.~K. {Effects of cationic starch addition and pulp beating on strength
  properties of softwood kraft pulp}. \emph{Starch - St{\"{a}}rke}
  \textbf{2014}, \emph{66}, 655--659\relax
\mciteBstWouldAddEndPuncttrue
\mciteSetBstMidEndSepPunct{\mcitedefaultmidpunct}
{\mcitedefaultendpunct}{\mcitedefaultseppunct}\relax
\EndOfBibitem
\bibitem[Lindstr{\"{o}}m \latin{et~al.}(2005)Lindstr{\"{o}}m, W\r{a}gberg, and
  Larsson]{Lindstrom2005}
Lindstr{\"{o}}m,~T.; W\r{a}gberg,~L.; Larsson,~T. {On the nature of joint
  strength in paper – a review of dry and wet strength resins used in paper
  manufacturing}. \emph{13th Fundamental Research Symposium} \textbf{2005},
  \relax
\mciteBstWouldAddEndPunctfalse
\mciteSetBstMidEndSepPunct{\mcitedefaultmidpunct}
{}{\mcitedefaultseppunct}\relax
\EndOfBibitem
\bibitem[Rojas and Azevedo(2011)Rojas, and Azevedo]{Rojas2011}
Rojas,~J.; Azevedo,~E. {Functionalization and crosslinking of microcrystalline
  cellulose in aqueous media: A safe and economic approach}. 2011\relax
\mciteBstWouldAddEndPuncttrue
\mciteSetBstMidEndSepPunct{\mcitedefaultmidpunct}
{\mcitedefaultendpunct}{\mcitedefaultseppunct}\relax
\EndOfBibitem
\bibitem[Xu \latin{et~al.}(2004)Xu, Yang, and Deng]{Xu2004}
Xu,~G.~G.; Yang,~C.~Q.; Deng,~Y. {Combination of bifunctional aldehydes and
  poly(vinyl alcohol) as the crosslinking systems to improve paper wet
  strength}. \emph{Journal of Applied Polymer Science} \textbf{2004},
  \emph{93}, 1673--1680\relax
\mciteBstWouldAddEndPuncttrue
\mciteSetBstMidEndSepPunct{\mcitedefaultmidpunct}
{\mcitedefaultendpunct}{\mcitedefaultseppunct}\relax
\EndOfBibitem
\bibitem[Gunnars \latin{et~al.}(2002)Gunnars, W\r{a}gberg, and {Cohen
  Stuart}]{Gunnars2002}
Gunnars,~S.; W\r{a}gberg,~L.; {Cohen Stuart},~M.~A. {Model films of cellulose:
  I. Method development and initial results}. \emph{Cellulose} \textbf{2002},
  \relax
\mciteBstWouldAddEndPunctfalse
\mciteSetBstMidEndSepPunct{\mcitedefaultmidpunct}
{}{\mcitedefaultseppunct}\relax
\EndOfBibitem
\bibitem[Kontturi \latin{et~al.}(2006)Kontturi, Tammelin, and
  {\"{O}}sterberg]{Kontturi2006}
Kontturi,~E.; Tammelin,~T.; {\"{O}}sterberg,~M. {Cellulose—model films and
  the fundamental approach}. \emph{Chemical Society Reviews} \textbf{2006},
  \relax
\mciteBstWouldAddEndPunctfalse
\mciteSetBstMidEndSepPunct{\mcitedefaultmidpunct}
{}{\mcitedefaultseppunct}\relax
\EndOfBibitem
\bibitem[Kontturi and Spirk(2019)Kontturi, and Spirk]{Kontturi2019}
Kontturi,~E.; Spirk,~S. {Ultrathin films of cellulose: A materials
  perspective}. 2019\relax
\mciteBstWouldAddEndPuncttrue
\mciteSetBstMidEndSepPunct{\mcitedefaultmidpunct}
{\mcitedefaultendpunct}{\mcitedefaultseppunct}\relax
\EndOfBibitem
\bibitem[Bismarck \latin{et~al.}(2002)Bismarck, Aranberri-Askargorta, Springer,
  Lampke, Wielage, Stamboulis, Shenderovich, and Limbach]{Bismarck2002}
Bismarck,~A.; Aranberri-Askargorta,~I.; Springer,~J.; Lampke,~T.; Wielage,~B.;
  Stamboulis,~A.; Shenderovich,~I.; Limbach,~H.~H. {Surface characterization of
  flax, hemp and cellulose fibers; Surface properties and the water uptake
  behavior}. \emph{Polymer Composites} \textbf{2002}, \relax
\mciteBstWouldAddEndPunctfalse
\mciteSetBstMidEndSepPunct{\mcitedefaultmidpunct}
{}{\mcitedefaultseppunct}\relax
\EndOfBibitem
\bibitem[Chinga-Carrasco(2009)]{Chinga2009}
Chinga-Carrasco,~G. {Exploring the multi-scale structure of printing paper - A
  review of modern technology}. \emph{Journal of Microscopy} \textbf{2009},
  \relax
\mciteBstWouldAddEndPunctfalse
\mciteSetBstMidEndSepPunct{\mcitedefaultmidpunct}
{}{\mcitedefaultseppunct}\relax
\EndOfBibitem
\bibitem[Medronho \latin{et~al.}(2012)Medronho, Romano, Miguel, Stigsson, and
  Lindman]{Medronho2012}
Medronho,~B.; Romano,~A.; Miguel,~M.~G.; Stigsson,~L.; Lindman,~B.
  {Rationalizing cellulose (in)solubility: reviewing basic physicochemical
  aspects and role of hydrophobic interactions}. \emph{Cellulose}
  \textbf{2012}, \emph{19}, 581--587\relax
\mciteBstWouldAddEndPuncttrue
\mciteSetBstMidEndSepPunct{\mcitedefaultmidpunct}
{\mcitedefaultendpunct}{\mcitedefaultseppunct}\relax
\EndOfBibitem
\bibitem[Dawsey and McCormick(1990)Dawsey, and McCormick]{Dawsey1990}
Dawsey,~T.~R.; McCormick,~C.~L. \emph{Journal of Macromolecular Science, Part
  C}; 1990; Vol.~30; pp 405--440\relax
\mciteBstWouldAddEndPuncttrue
\mciteSetBstMidEndSepPunct{\mcitedefaultmidpunct}
{\mcitedefaultendpunct}{\mcitedefaultseppunct}\relax
\EndOfBibitem
\bibitem[Kargl \latin{et~al.}(2015)Kargl, Mohan, Ribitsch, Saake, Puls, and
  Stana-Kleinschek]{Kargl2015}
Kargl,~R.; Mohan,~T.; Ribitsch,~V.; Saake,~B.; Puls,~J.; Stana-Kleinschek,~K.
  {Cellulose thin films from ionic liquid solutions}. \emph{Nordic Pulp and
  Paper Research Journal} \textbf{2015}, \relax
\mciteBstWouldAddEndPunctfalse
\mciteSetBstMidEndSepPunct{\mcitedefaultmidpunct}
{}{\mcitedefaultseppunct}\relax
\EndOfBibitem
\bibitem[Edgar and Gray(2003)Edgar, and Gray]{Edgar2003}
Edgar,~C.~D.; Gray,~D.~G. {Smooth model cellulose I surfaces from nanocrystal
  suspensions}. \emph{Cellulose} \textbf{2003}, \relax
\mciteBstWouldAddEndPunctfalse
\mciteSetBstMidEndSepPunct{\mcitedefaultmidpunct}
{}{\mcitedefaultseppunct}\relax
\EndOfBibitem
\bibitem[Aulin \latin{et~al.}(2009)Aulin, Ahok, Josefsson, Nishino, Hirose,
  {\"{O}}sterberg, and W{\aa}gberg]{Aulin2009}
Aulin,~C.; Ahok,~S.; Josefsson,~P.; Nishino,~T.; Hirose,~Y.;
  {\"{O}}sterberg,~M.; W{\aa}gberg,~L. {Nanoscale cellulose films with
  different crystallinities and mesostructures - Their surface properties and
  interaction with water}. \emph{Langmuir} \textbf{2009}, \relax
\mciteBstWouldAddEndPunctfalse
\mciteSetBstMidEndSepPunct{\mcitedefaultmidpunct}
{}{\mcitedefaultseppunct}\relax
\EndOfBibitem
\bibitem[Jahan \latin{et~al.}(2009)Jahan, Noori, Ahsan, Nasima, and
  Quaiyyum]{Jahan2009}
Jahan,~M.~S.; Noori,~A.; Ahsan,~L.; Nasima,~C.~D.; Quaiyyum,~M.~A. {Effects of
  chitosan as dry and wet strength additive in bamboo and acacia pulp}.
  \emph{IPPTA: Quarterly Journal of Indian Pulp and Paper Technical
  Association} \textbf{2009}, \relax
\mciteBstWouldAddEndPunctfalse
\mciteSetBstMidEndSepPunct{\mcitedefaultmidpunct}
{}{\mcitedefaultseppunct}\relax
\EndOfBibitem
\bibitem[Taketa \latin{et~al.}(2018)Taketa, {Dos Santos}, Fiamingo, Vaz, Beppu,
  Campana-Filho, Cohen, and Rubner]{Taketa2018}
Taketa,~T.~B.; {Dos Santos},~D.~M.; Fiamingo,~A.; Vaz,~J.~M.; Beppu,~M.~M.;
  Campana-Filho,~S.~P.; Cohen,~R.~E.; Rubner,~M.~F. {Investigation of the
  Internal Chemical Composition of Chitosan-Based LbL Films by Depth-Profiling
  X-ray Photoelectron Spectroscopy (XPS) Analysis}. \emph{Langmuir}
  \textbf{2018}, \relax
\mciteBstWouldAddEndPunctfalse
\mciteSetBstMidEndSepPunct{\mcitedefaultmidpunct}
{}{\mcitedefaultseppunct}\relax
\EndOfBibitem
\bibitem[Bataglioli \latin{et~al.}(2019)Bataglioli, Taketa, Neto, Lopes, Costa,
  and Beppu]{Bataglioli2019}
Bataglioli,~R.~A.; Taketa,~T.~B.; Neto,~J.~B.; Lopes,~L.~M.; Costa,~C.~A.;
  Beppu,~M.~M. {Analysis of pH and salt concentration on structural and
  model-drug delivery properties of polysaccharide-based multilayered films}.
  \emph{Thin Solid Films} \textbf{2019}, \relax
\mciteBstWouldAddEndPunctfalse
\mciteSetBstMidEndSepPunct{\mcitedefaultmidpunct}
{}{\mcitedefaultseppunct}\relax
\EndOfBibitem
\bibitem[Spera \latin{et~al.}(2017)Spera, Taketa, and Beppu]{Spera2017}
Spera,~M. B.~M.; Taketa,~T.~B.; Beppu,~M.~M. {Roughness dynamic in surface
  growth: Layer-by-layer thin films of carboxymethyl cellulose/chitosan for
  biomedical applications}. \emph{Biointerphases} \textbf{2017}, \emph{12},
  04E401\relax
\mciteBstWouldAddEndPuncttrue
\mciteSetBstMidEndSepPunct{\mcitedefaultmidpunct}
{\mcitedefaultendpunct}{\mcitedefaultseppunct}\relax
\EndOfBibitem
\bibitem[Zhang \latin{et~al.}(2013)Zhang, Liu, Liang, Li, Liang, He, Zhu, and
  Mao]{Zhang2013}
Zhang,~S.; Liu,~W.; Liang,~J.; Li,~X.; Liang,~W.; He,~S.; Zhu,~C.; Mao,~L.
  {Buildup mechanism of carboxymethyl cellulose and chitosan self-assembled
  films}. \emph{Cellulose} \textbf{2013}, \relax
\mciteBstWouldAddEndPunctfalse
\mciteSetBstMidEndSepPunct{\mcitedefaultmidpunct}
{}{\mcitedefaultseppunct}\relax
\EndOfBibitem
\bibitem[L{\"{o}}hmann \latin{et~al.}(2018)L{\"{o}}hmann, Zerball, and {Von
  Klitzing}]{Lohmann2018}
L{\"{o}}hmann,~O.; Zerball,~M.; {Von Klitzing},~R. {Water Uptake of
  Polyelectrolyte Multilayers Including Water Condensation in Voids}.
  \emph{Langmuir} \textbf{2018}, \relax
\mciteBstWouldAddEndPunctfalse
\mciteSetBstMidEndSepPunct{\mcitedefaultmidpunct}
{}{\mcitedefaultseppunct}\relax
\EndOfBibitem
\bibitem[Lavalle \latin{et~al.}(2004)Lavalle, Picart, Mutterer, Gergely, Reiss,
  Voegel, Senger, and Schaaf]{Lavalle2004}
Lavalle,~P.; Picart,~C.; Mutterer,~J.; Gergely,~C.; Reiss,~H.; Voegel,~J.~C.;
  Senger,~B.; Schaaf,~P. {Modeling the Buildup of Polyelectrolyte Multilayer
  Films Having Exponential Growth}. \emph{Journal of Physical Chemistry B}
  \textbf{2004}, \relax
\mciteBstWouldAddEndPunctfalse
\mciteSetBstMidEndSepPunct{\mcitedefaultmidpunct}
{}{\mcitedefaultseppunct}\relax
\EndOfBibitem
\bibitem[Tang and Besseling(2016)Tang, and Besseling]{Tang2016}
Tang,~K.; Besseling,~N. A.~M. {Formation of polyelectrolyte multilayers: ionic
  strengths and growth regimes}. \emph{Soft Matter} \textbf{2016}, \emph{12},
  1032--1040\relax
\mciteBstWouldAddEndPuncttrue
\mciteSetBstMidEndSepPunct{\mcitedefaultmidpunct}
{\mcitedefaultendpunct}{\mcitedefaultseppunct}\relax
\EndOfBibitem
\bibitem[Volodkin and von Klitzing(2014)Volodkin, and von
  Klitzing]{Volodkin2014}
Volodkin,~D.; von Klitzing,~R. {Competing mechanisms in polyelectrolyte
  multilayer formation and swelling: Polycation–polyanion pairing vs.
  polyelectrolyte–ion pairing}. \emph{Current Opinion in Colloid \& Interface
  Science} \textbf{2014}, \emph{19}, 25--31\relax
\mciteBstWouldAddEndPuncttrue
\mciteSetBstMidEndSepPunct{\mcitedefaultmidpunct}
{\mcitedefaultendpunct}{\mcitedefaultseppunct}\relax
\EndOfBibitem
\bibitem[Kujawa \latin{et~al.}(2005)Kujawa, Moraille, Sanchez, Badia, and
  Winnik]{Kujawa2005}
Kujawa,~P.; Moraille,~P.; Sanchez,~J.; Badia,~A.; Winnik,~F.~M. {Effect of
  Molecular Weight on the Exponential Growth and Morphology of
  Hyaluronan/Chitosan Multilayers: A Surface Plasmon Resonance Spectroscopy and
  Atomic Force Microscopy Investigation}. \emph{Journal of the American
  Chemical Society} \textbf{2005}, \emph{127}, 9224--9234\relax
\mciteBstWouldAddEndPuncttrue
\mciteSetBstMidEndSepPunct{\mcitedefaultmidpunct}
{\mcitedefaultendpunct}{\mcitedefaultseppunct}\relax
\EndOfBibitem
\bibitem[Abdelkebir \latin{et~al.}(2011)Abdelkebir, Gaudi{\`{e}}re,
  Morin-Grognet, Coquerel, Labat, Atmani, and Ladam]{Abdelkebir2011}
Abdelkebir,~K.; Gaudi{\`{e}}re,~F.; Morin-Grognet,~S.; Coquerel,~G.; Labat,~B.;
  Atmani,~H.; Ladam,~G. {Evidence of different growth regimes coexisting within
  biomimetic Layer-by-Layer films}. \emph{Soft Matter} \textbf{2011}, \relax
\mciteBstWouldAddEndPunctfalse
\mciteSetBstMidEndSepPunct{\mcitedefaultmidpunct}
{}{\mcitedefaultseppunct}\relax
\EndOfBibitem
\bibitem[Porcel \latin{et~al.}(2006)Porcel, Lavalle, Ball, Decher, Senger,
  Voegel, and Schaaf]{Porcel2006}
Porcel,~C.; Lavalle,~P.; Ball,~V.; Decher,~G.; Senger,~B.; Voegel,~J.~C.;
  Schaaf,~P. {From Exponential to Linear Growth in Polyelectrolyte
  Multilayers}. \emph{Langmuir} \textbf{2006}, \relax
\mciteBstWouldAddEndPunctfalse
\mciteSetBstMidEndSepPunct{\mcitedefaultmidpunct}
{}{\mcitedefaultseppunct}\relax
\EndOfBibitem
\bibitem[Wang \latin{et~al.}(2016)Wang, Xu, Backes, Li, Micciulla, Kayitmazer,
  Li, Guo, and von Klitzing]{Wang2016}
Wang,~W.; Xu,~Y.; Backes,~S.; Li,~A.; Micciulla,~S.; Kayitmazer,~A.~B.; Li,~L.;
  Guo,~X.; von Klitzing,~R. {Construction of Compact Polyelectrolyte
  Multilayers Inspired by Marine Mussel: Effects of Salt Concentration and pH
  As Observed by QCM-D and AFM}. \emph{Langmuir} \textbf{2016}, \emph{32},
  3365--3374\relax
\mciteBstWouldAddEndPuncttrue
\mciteSetBstMidEndSepPunct{\mcitedefaultmidpunct}
{\mcitedefaultendpunct}{\mcitedefaultseppunct}\relax
\EndOfBibitem
\bibitem[von Klitzing(2006)]{V.Klitzing2006}
von Klitzing,~R. {Internal structure of polyelectrolyte multilayer assemblies}.
  2006\relax
\mciteBstWouldAddEndPuncttrue
\mciteSetBstMidEndSepPunct{\mcitedefaultmidpunct}
{\mcitedefaultendpunct}{\mcitedefaultseppunct}\relax
\EndOfBibitem
\bibitem[Steitz \latin{et~al.}(2001)Steitz, Jaeger, and von
  Klitzing]{Steitz2001}
Steitz,~R.; Jaeger,~W.; von Klitzing,~R. {Influence of Charge Density and Ionic
  Strength on the Multilayer Formation of Strong Polyelectrolytes}.
  \emph{Langmuir} \textbf{2001}, \emph{17}, 4471--4474\relax
\mciteBstWouldAddEndPuncttrue
\mciteSetBstMidEndSepPunct{\mcitedefaultmidpunct}
{\mcitedefaultendpunct}{\mcitedefaultseppunct}\relax
\EndOfBibitem
\bibitem[Dubas and Schlenoff(1999)Dubas, and Schlenoff]{Dubas1999}
Dubas,~S.~T.; Schlenoff,~J.~B. {Factors Controlling the Growth of
  Polyelectrolyte Multilayers}. \emph{Macromolecules} \textbf{1999}, \emph{32},
  8153--8160\relax
\mciteBstWouldAddEndPuncttrue
\mciteSetBstMidEndSepPunct{\mcitedefaultmidpunct}
{\mcitedefaultendpunct}{\mcitedefaultseppunct}\relax
\EndOfBibitem
\bibitem[Garg \latin{et~al.}(2008)Garg, Heflin, Gibson, and Davis]{Garg2008}
Garg,~A.; Heflin,~J.~R.; Gibson,~H.~W.; Davis,~R.~M. {Study of Film Structure
  and Adsorption Kinetics of Polyelectrolyte Multilayer Films: Effect of pH and
  Polymer Concentration}. \emph{Langmuir} \textbf{2008}, \emph{24},
  10887--10894\relax
\mciteBstWouldAddEndPuncttrue
\mciteSetBstMidEndSepPunct{\mcitedefaultmidpunct}
{\mcitedefaultendpunct}{\mcitedefaultseppunct}\relax
\EndOfBibitem
\bibitem[Mermut and Barrett(2003)Mermut, and Barrett]{Mermut2003}
Mermut,~O.; Barrett,~C.~J. {Effects of Charge Density and Counterions on the
  Assembly of Polyelectrolyte Multilayers}. \emph{The Journal of Physical
  Chemistry B} \textbf{2003}, \emph{107}, 2525--2530\relax
\mciteBstWouldAddEndPuncttrue
\mciteSetBstMidEndSepPunct{\mcitedefaultmidpunct}
{\mcitedefaultendpunct}{\mcitedefaultseppunct}\relax
\EndOfBibitem
\bibitem[Sui \latin{et~al.}(2003)Sui, Salloum, and Schlenoff]{Sui2003}
Sui,~Z.; Salloum,~D.; Schlenoff,~J.~B. {Effect of Molecular Weight on the
  Construction of Polyelectrolyte Multilayers: Stripping versus Sticking}.
  \emph{Langmuir} \textbf{2003}, \emph{19}, 2491--2495\relax
\mciteBstWouldAddEndPuncttrue
\mciteSetBstMidEndSepPunct{\mcitedefaultmidpunct}
{\mcitedefaultendpunct}{\mcitedefaultseppunct}\relax
\EndOfBibitem
\bibitem[Fery \latin{et~al.}(2001)Fery, Sch{\"{o}}ler, Cassagneau, and
  Caruso]{Fery2001}
Fery,~A.; Sch{\"{o}}ler,~B.; Cassagneau,~T.; Caruso,~F. {Nanoporous thin films
  formed by salt-induced structural changes in multilayers of poly(acrylic
  acid) and poly(allylamine)}. \emph{Langmuir} \textbf{2001}, \relax
\mciteBstWouldAddEndPunctfalse
\mciteSetBstMidEndSepPunct{\mcitedefaultmidpunct}
{}{\mcitedefaultseppunct}\relax
\EndOfBibitem
\bibitem[Lehaf \latin{et~al.}(2012)Lehaf, Hariri, and Schlenoff]{Lehaf2012}
Lehaf,~A.~M.; Hariri,~H.~H.; Schlenoff,~J.~B. {Homogeneity, modulus, and
  viscoelasticity of polyelectrolyte multilayers by nanoindentation: Refining
  the buildup mechanism}. \emph{Langmuir} \textbf{2012}, \relax
\mciteBstWouldAddEndPunctfalse
\mciteSetBstMidEndSepPunct{\mcitedefaultmidpunct}
{}{\mcitedefaultseppunct}\relax
\EndOfBibitem
\bibitem[Fenstermaker and McCrackin(1969)Fenstermaker, and
  McCrackin]{Fenstermaker1969}
Fenstermaker,~C.~A.; McCrackin,~F.~L. {Errors arising from surface roughness in
  ellipsometric measurement of the refractive index of a surface}.
  \emph{Surface Science} \textbf{1969}, \emph{16}, 85--96\relax
\mciteBstWouldAddEndPuncttrue
\mciteSetBstMidEndSepPunct{\mcitedefaultmidpunct}
{\mcitedefaultendpunct}{\mcitedefaultseppunct}\relax
\EndOfBibitem
\bibitem[Meier \latin{et~al.}(2011)Meier, Ruderer, Diethert, Kaune, K"orstgens,
  Roth, and M"uller-Buschbaum]{Meier2011}
Meier,~R.; Ruderer,~M.~A.; Diethert,~A.; Kaune,~G.; K"orstgens,~V.;
  Roth,~S.~V.; M"uller-Buschbaum,~P. {Influence of Film Thickness on the Phase
  Separation Mechanism in Ultrathin Conducting Polymer Blend Films}. \emph{The
  Journal of Physical Chemistry B} \textbf{2011}, \emph{115}, 2899--2909\relax
\mciteBstWouldAddEndPuncttrue
\mciteSetBstMidEndSepPunct{\mcitedefaultmidpunct}
{\mcitedefaultendpunct}{\mcitedefaultseppunct}\relax
\EndOfBibitem
\bibitem[Ghoussoub \latin{et~al.}(2018)Ghoussoub, Zerball, Fares, Ankner, von
  Klitzing, and Schlenoff]{Ghoussoub2018}
Ghoussoub,~Y.~E.; Zerball,~M.; Fares,~H.~M.; Ankner,~J.~F.; von Klitzing,~R.;
  Schlenoff,~J.~B. {Ion distribution in dry polyelectrolyte multilayers: a
  neutron reflectometry study}. \emph{Soft Matter} \textbf{2018}, \emph{14},
  1699--1708\relax
\mciteBstWouldAddEndPuncttrue
\mciteSetBstMidEndSepPunct{\mcitedefaultmidpunct}
{\mcitedefaultendpunct}{\mcitedefaultseppunct}\relax
\EndOfBibitem
\bibitem[Wong \latin{et~al.}(2004)Wong, Rehfeldt, H{\"{a}}nni, Tanaka, and von
  Klitzing]{Wong2004}
Wong,~J.~E.; Rehfeldt,~F.; H{\"{a}}nni,~P.; Tanaka,~M.; von Klitzing,~R.
  {Swelling Behavior of Polyelectrolyte Multilayers in Saturated Water Vapor}.
  \emph{Macromolecules} \textbf{2004}, \emph{37}, 7285--7289\relax
\mciteBstWouldAddEndPuncttrue
\mciteSetBstMidEndSepPunct{\mcitedefaultmidpunct}
{\mcitedefaultendpunct}{\mcitedefaultseppunct}\relax
\EndOfBibitem
\bibitem[Wang \latin{et~al.}(2017)Wang, Xu, Han, Micciulla, Backes, Li, Xu,
  Shen, von Klitzing, and Guo]{Wang2017}
Wang,~W.; Xu,~Y.; Han,~H.; Micciulla,~S.; Backes,~S.; Li,~A.; Xu,~J.; Shen,~W.;
  von Klitzing,~R.; Guo,~X. {Odd-even effect during layer-by-layer assembly of
  polyelectrolytes inspired by marine mussel}. \emph{Journal of Polymer Science
  Part B: Polymer Physics} \textbf{2017}, \emph{55}, 245--255\relax
\mciteBstWouldAddEndPuncttrue
\mciteSetBstMidEndSepPunct{\mcitedefaultmidpunct}
{\mcitedefaultendpunct}{\mcitedefaultseppunct}\relax
\EndOfBibitem
\bibitem[Zerball \latin{et~al.}(2015)Zerball, Laschewsky, and von
  Klitzing]{Zerball2015}
Zerball,~M.; Laschewsky,~A.; von Klitzing,~R. {Swelling of Polyelectrolyte
  Multilayers: The Relation Between, Surface and Bulk Characteristics}.
  \emph{The Journal of Physical Chemistry B} \textbf{2015}, \emph{119},
  11879--11886\relax
\mciteBstWouldAddEndPuncttrue
\mciteSetBstMidEndSepPunct{\mcitedefaultmidpunct}
{\mcitedefaultendpunct}{\mcitedefaultseppunct}\relax
\EndOfBibitem
\bibitem[Nestler \latin{et~al.}(2012)Nestler, Block, and Helm]{Nestler2012}
Nestler,~P.; Block,~S.; Helm,~C.~A. {Temperature-Induced Transition from
  Odd–Even to Even–Odd Effect in Polyelectrolyte Multilayers Due to
  Interpolyelectrolyte Interactions}. \emph{The Journal of Physical Chemistry
  B} \textbf{2012}, \emph{116}, 1234--1243\relax
\mciteBstWouldAddEndPuncttrue
\mciteSetBstMidEndSepPunct{\mcitedefaultmidpunct}
{\mcitedefaultendpunct}{\mcitedefaultseppunct}\relax
\EndOfBibitem
\bibitem[Lavalle \latin{et~al.}(2002)Lavalle, Gergely, Cuisinier, Decher,
  Schaaf, Voegel, and Picart]{Lavalle2002}
Lavalle,~P.; Gergely,~C.; Cuisinier,~F.~J.; Decher,~G.; Schaaf,~P.;
  Voegel,~J.~C.; Picart,~C. {Comparison of the structure of polyelectrolyte
  multilayer films exhibiting a linear and an exponential growth regime: An in
  situ atomic force microscopy study}. \emph{Macromolecules} \textbf{2002},
  \relax
\mciteBstWouldAddEndPunctfalse
\mciteSetBstMidEndSepPunct{\mcitedefaultmidpunct}
{}{\mcitedefaultseppunct}\relax
\EndOfBibitem
\end{mcitethebibliography}

\end{document}